%% file: main.tex
\documentclass[12pt]{iopart}

\usepackage{acronym}
\usepackage{cite}
\usepackage{orcidlink}
\definecolor{orcidlogocol}{HTML}{A6CE39}    
\usepackage{csquotes}
\usepackage{subfigure}
\usepackage{amssymb,mathrsfs}
\usepackage{siunitx}
\usepackage{helvet}
\usepackage{hyperref}
\hypersetup{
    colorlinks=true,
    breaklinks=true,
    citecolor=black,
    linkcolor=black,
    menucolor=black,
    urlcolor=black,
}
\begin{document}
\paper[Testing Inertial Sensors in a Gravitational Wave Detector Prototype]{Testing Compact, Fused Silica Resonator Based Inertial Sensors in a Gravitational Wave Detector Prototype Facility}

%\linenumbers
\author{J J Carter$^{1,2}$\orcidlink{0000-0001-8845-0900}, P Birckigt$^{3}$\orcidlink{0000-0001-8492-5964}, J Lehmann$^{1,2}$\orcidlink{0009-0000-6998-4413}, A Basalaev$^{1,2}$\orcidlink{0000-0001-5623-2853}, S L Kranzhoff$^{4,5}$\orcidlink{0000-0003-3533-2059}, S Al-Kershi$^{1,2}$\orcidlink{ 0009-0004-1844-2381}, M Carlassara$^{1,2}$\orcidlink{0009-0007-2345-3706}, G Chiarini$^{1,2}$\orcidlink{0009-0001-2977-5825}, F Khan $^{1,2}$\orcidlink{0000-0001-6176-853X}, G Leibeling$^3$, H Lück$^{1,2}$\orcidlink{0000-0001-9350-4846}, C Rothhardt$^3$\orcidlink{0000-0002-2932-6165}, S Risse$^3$, P Sarkar$^{1,2}$\orcidlink{0009-0009-4054-6888}, S Takano$^{1,2}$\orcidlink{0000-0002-1266-4555}, J von Wrangel$^{1,2}$\orcidlink{0009-0006-8431-0533}, D S Wu$^{1,2}$\orcidlink{0000-0003-2849-3751} and S M Koehlenbeck$^{6}$\orcidlink{0000-0002-3842-9051}}
\address{$^1$ Max Planck Institute for Gravitational Physics (Albert Einstein Institute), Callinstr. 38, Hannover, Germany}
\address{$^2$ Institute for Gravitational Physics of the Leibniz Universit\"at Hannover, Callinstr. 38, Hannover, Germany}
\address{$^3$ Fraunhofer Institute for Applied Optics and Precision Engineering, Albert-Einstein-Str.~7, Jena, Germany}
\address{$^4$ Universiteit Maastricht, Maastricht, 6200 MD, The Netherlands}
\address{$^5$ Nikhef, Science Park 105, Amsterdam, 1098 XG, The Netherlands}
\address{$^6$ E.\ L.\ Ginzton Laboratory, Stanford University, 348 Via Pueblo, Stanford, CA 94305, USA}
\ead{Jonathan.carter@aei.mpg.de}
\date{\today}

\begin{abstract}
Future gravitational wave observatories require significant advances in all aspects of their seismic isolation; inertial sensors being a pressing example. Inertial sensors using gram-scale high mechanical Q factor (Q) glass resonators combined with compact interferometric readout are promising alternatives to kilogram-scale conventional inertial sensors. We have produced fused silica resonators suitable for low frequency inertial sensing and demonstrated that Qs of over 150,000 are possible. One resonator we produced was combined with a homodyne quadrature interferometer (HoQI) to read out the test mass displacement to form an inertial sensor. This is the first time a HoQI was used with a high Q resonator. The resulting sensor was tested against other commercial, kilogram scale inertial sensors at the AEI 10\,m Prototype facility. Despite the dynamic range challenges induced by the test mass motion, we can match the excellent noise floors HoQIs have achieved so far with slow-moving or stationary test masses, showing HoQIs as an excellent candidate for the readout of such sensors. We evaluate the setup as an inertial sensor, showing the best performance demonstrated by any gram-scale sensor to date, with comparable sensitivity to the significantly bulkier sensors used in gravitational wave detectors today. These sensors' compact size, self-calibration, and vacuum compatibility make them ideal candidates for the inertial sensing requirements in future gravitational wave detectors. %As an outlook, the scalability of the manufacturing method used for this resonator will be used to make much larger fused silica resonators, which will significantly improve future sensor sensitivity.
\end{abstract}

\maketitle
\noindent{\it Keywords\/}: Inertial sensing, Interferometric readout, Gravitational wave detectors 
\section{Introduction}
Terrestrial gravitational wave detectors require sophisticated technologies to sufficiently isolate their mirrors, which act as test masses, from seismic motion~\cite{Matichard2015a,Matichard2015b}. To do this, we must measure ground motion to exceptional precision in the control band of 10\,mHz to 30\,Hz. This requires a suite of inertial and displacement sensors with incredibly low self-noise. Despite our best efforts, seismic and environmental disturbances can cause lock loss, so there are concerted efforts to tackle all aspects of low-frequency control noise~\cite{Capote2025,Acernese2023,Koehlenbeck2023}. One of the pillars of this effort is the development of better inertial sensors for local control of the test masses~\cite{Heijningen2022,Cooper2022,Kirchhoff2020,Kranzhoff2022,Hines2023,Carter2024b}.\par
Gram-scale, Fused Silica resonator based Inertial Sensors  (FuSIS) are a technology seeing rapid development~\cite{Guzman2014,Carter2020a,Hines2020,Hines2023,Capistran2023,Carter2024b}. These sensors typically use the high mechanical Quality factor (Q) facilitated by  the well-defined behaviour of fused silica. They are compact packages for applications where extreme precision is needed alongside assembly space, weight or vacuum compatibility requirements; for example, in the seismic isolation systems of gravitational wave detectors~\cite{Hines2023,Carter2024b}, in space-based control of satellites~\cite{Sanjuan2022,Sanjuan2023}, and in atomic interferometry experiments~\cite{Richardson2020}.
The field is now maturing, and recent research has focused on understanding the behaviour of the mechanical resonators used in such sensors~\cite{Kumanchik2023,Carter2024,Rezinkina2024}, novel manufacturing methods for the necessary resonators~\cite{Guzman2014,Bellouard2012, Nelson2022,Birckigt2024}, and development of novel readout techniques of test mass displacement in compact sensors~\cite{Hines2023,Carter2024b,Yang2020,Zhou2021,Zhang2021,ChalathadkaSubrahmanya2025}. The test mass displacement readout of a high Q resonator is a difficult task as on resonance, the motion of the test mass is substantial, ranging from nanometre to micrometre scales based on resonance frequency and environment, with a precision requirement to remain competitive typically of sub picometre. 
\par
A novel, scalable method of producing gram scale resonators was recently published~\cite{Birckigt2024}. The technique demonstrated that Qs above 100,000 can be achieved with this design method. With these promising initial results, we wished to use them in a FuSIS.  We, therefore, combined one of the resonators produced via this method with a suitable means of measuring the displacement of its test mass.
\par
Homodyne Quadrature Interferometers (HoQIs) have been a tool for measuring the displacement of test masses in the gravitational wave community for several years now~\cite{Cooper2018,Ross2023}. The technique uses orthogonal polarisation states of light to convert a probe of a test and reference mass into three signals. Combinations of these signals can then give a displacement readout of the test mass that is not susceptible to the coupling of the intensity noise of the laser. All this can be done in compact packages typically of several centimetres cubed~\cite{Kranzhoff2022}. HoQIs have been used to probe the motion of kilogram scale inertial sensors and have shown that they can reach noise floors that can be fully characterised~\cite{Kranzhoff2022,Cooper2022,Ross2023}. 
\par 
We combine a high Q factor, fused silica mechanical resonator~\cite{Birckigt2024}, with a HoQI to create a FuSIS. We characterise the FuSIS in a huddle test, a technique to subtract coherent signals from witness sensors and reveal the sensor noise. We do this test in the Albert Einstein Institute in Hannover's (AEI) 10 m Prototype Facility for gravitational wave detector technology~\cite{Westphal2012,Lehmann2024} using several state-of-the-art inertial sensors already installed there~\cite{Bertolini2006,Kirchhoff2017}. We show that HoQIs can track the large dynamic range needed in a FuSIS without loss in performance.  As an inertial sensor, the FuSIS's self-noise is, to our knowledge, the lowest of a fused silica, high Q design in the literature to date. %(Alternative, shorter sentence: As an inertial sensor, the self-noise is, to our knowledge, the lowest of any FuSIS to date.)

\section{Components and Implementation of the Inertial Sensor}
Any inertial sensor needs two things: a means of encoding forces and a means of reading out that encoded force. Whichever of these two performs worse will limit the sensor's performance. Most inertial sensors use a suspended mass for the former, encoding forces into motion.  There are a plethora of methods of reading out the induced motion~\cite{Hines2023,Carter2024b,Yang2020,Zhou2021,Zhang2021,ChalathadkaSubrahmanya2025}.
Before presenting our specific experimental design, we motivate our decisions for both readout and resonator from basic physical principles.
\par 
\subsection{Theoretical basis of inertial sensor design}

The simplest method of encoding motion is to suspend a test mass using a soft spring. In a fused silica resonator, the soft spring is created using a thin bridge of fused silica typically around \SI{100}{\micro\meter} thick~\cite{Guzman2014,Gerberding2015}. This suspension creates the fundamental limit of an inertial sensor's performance by coupling it to suspension thermal noise. Through the fluctuation-dissipation theorem, the motion of particles in a heat bath is directly coupled to excitations of the mechanical modes of the resonator. The scale of this coupling for a structurally damped resonator is given by 
\begin{equation}
	\tilde{A}_{th}(f)=\sqrt{\frac{8\pi k_{\rm{b}}Tf_0^2}{ mf Q}}
	\label{eqn:strucNoise}
\end{equation}
where $\tilde{A}_{th}(f)$ is the spectral density of thermally induced acceleration noise, $k_{\rm{b}}$ is the Boltzmann constant, $T$ is the temperature, $f_0$ the natural frequency of the resonator, $f$ the Fourier frequency, $m$ the suspended mass, and $Q$ the mechanical Quality factor of the resonator. Most classical inertial sensors suppress this noise with a kilogram scale suspended mass~\cite{Cooper2022}, but fused silica resonators use the low loss of the material to utilise high Q for their resonant modes, significantly compacting the size of the sensor needed to achieve the necessary performance.
\par
Most fused silica resonator-based inertial sensors (FuSIS) use interferometric readout, as the pieces are typically made of optical-grade fused silica, allowing them to be locally coated and used as an end mirror of an interferometer \cite{Hines2023,Carter2024b,Zhou2021}. The noise floor of the readout is given by 
\begin{equation}
    {\tilde{A}_{\rm{rn}}}(f)=(2\pi)^2\left| f^2-{f_0}^2-\frac{if_0f}{Q}\right|{\Delta \tilde{Z}}
    \label{eqn:tfseismo}
\end{equation}
where $\tilde{A}_{\rm{rn}}$ is the acceleration noise floor from readout and $\Delta \tilde{Z}$ is the measured displacement noise floor of the readout~\cite{Carter2024}. Due to the extremely high Q that is possible when using fused silica, recent publications in this field have mostly been limited by readout noise \cite{Carter2024b,Hines2023}, but some have been thermal noise floor limited \cite{Zhou2021}. For the readout limited case, we see that the only two things that can significantly improve the overall noise floor are lowering the readout noise or the fundamental frequency of the resonator. This means the unparalleled sensitivity of interferometry is typically required to compensate for the higher $f_0$ than kilogram scale sensors. 
\par
The necessity and difficulties in lowering $f_0$ in FuSIS are discussed in detail in~\cite{Carter2024}. Lowering $f_0$ is realized by designing very thin flexure bridges, which makes them prone to brittle fracture, requiring longer bridges in turn. 
\subsection{The Fused Silica Mechanical Resonator}
\begin{figure}
    \centering
    \subfigure[Coloured Resonator Design]{\includegraphics[width=.3\textwidth]{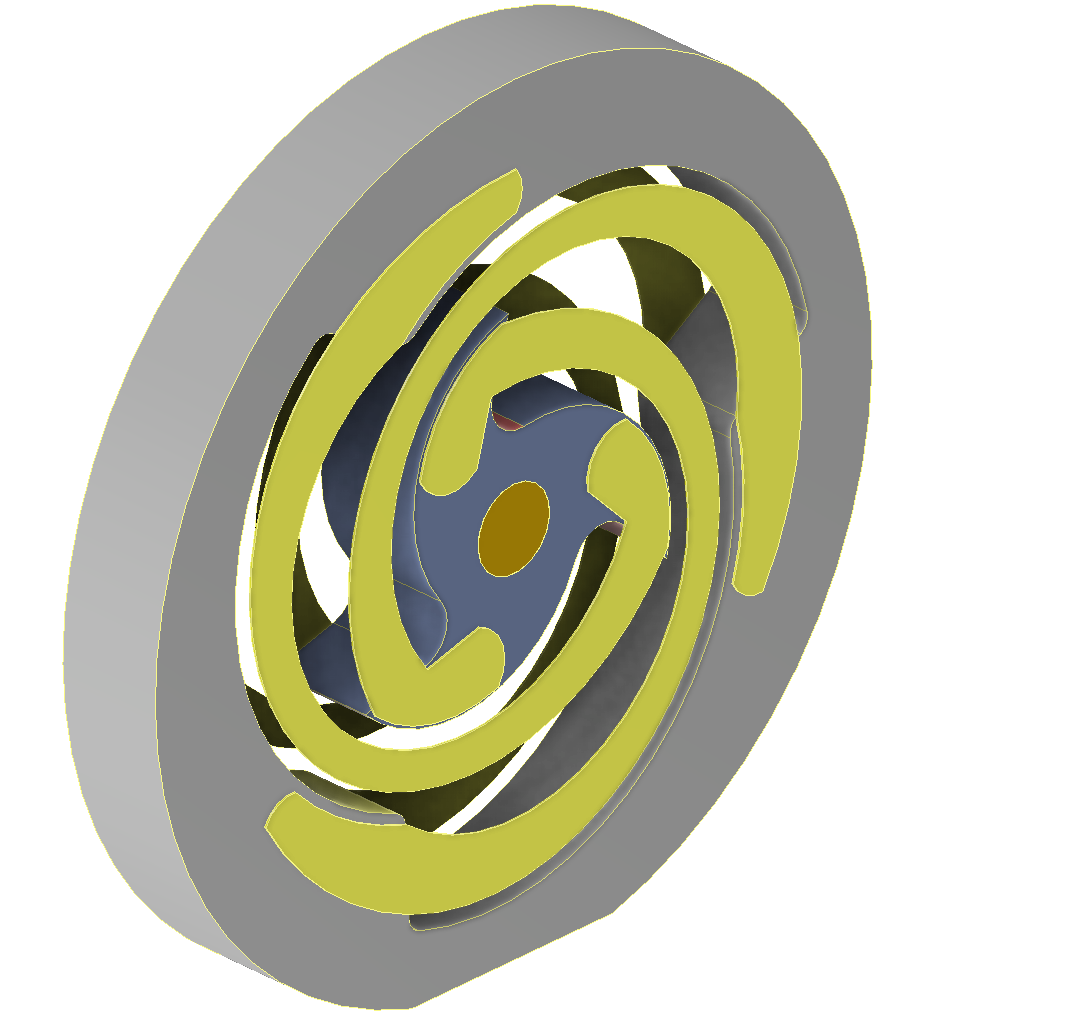}\hfill}
    \subfigure[Normalised Z Displacement from Excitation]{\includegraphics[width=.48\textwidth]{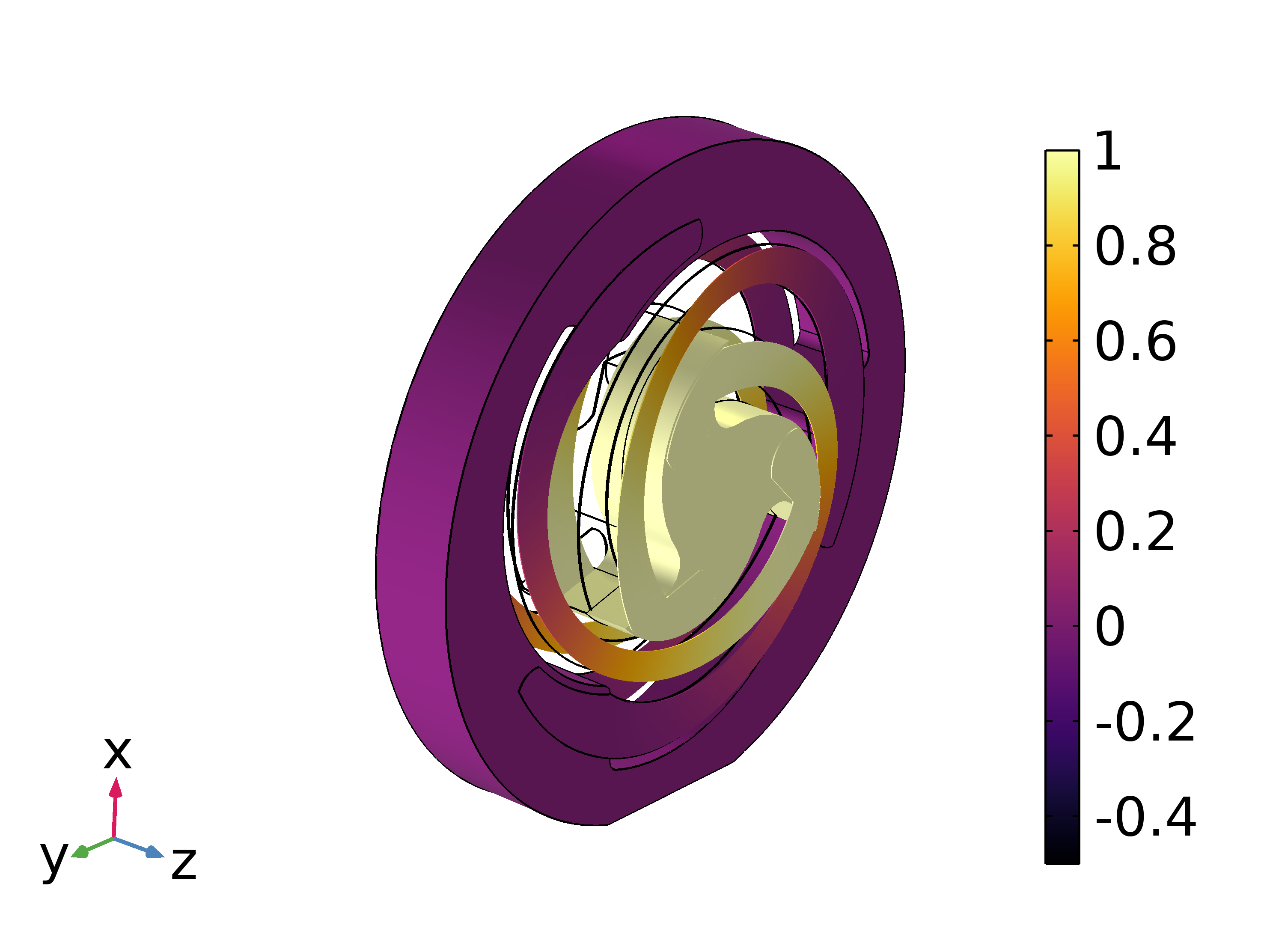}\hfill}
    \subfigure[Mount Photo]{\includegraphics[width=.3\textwidth]{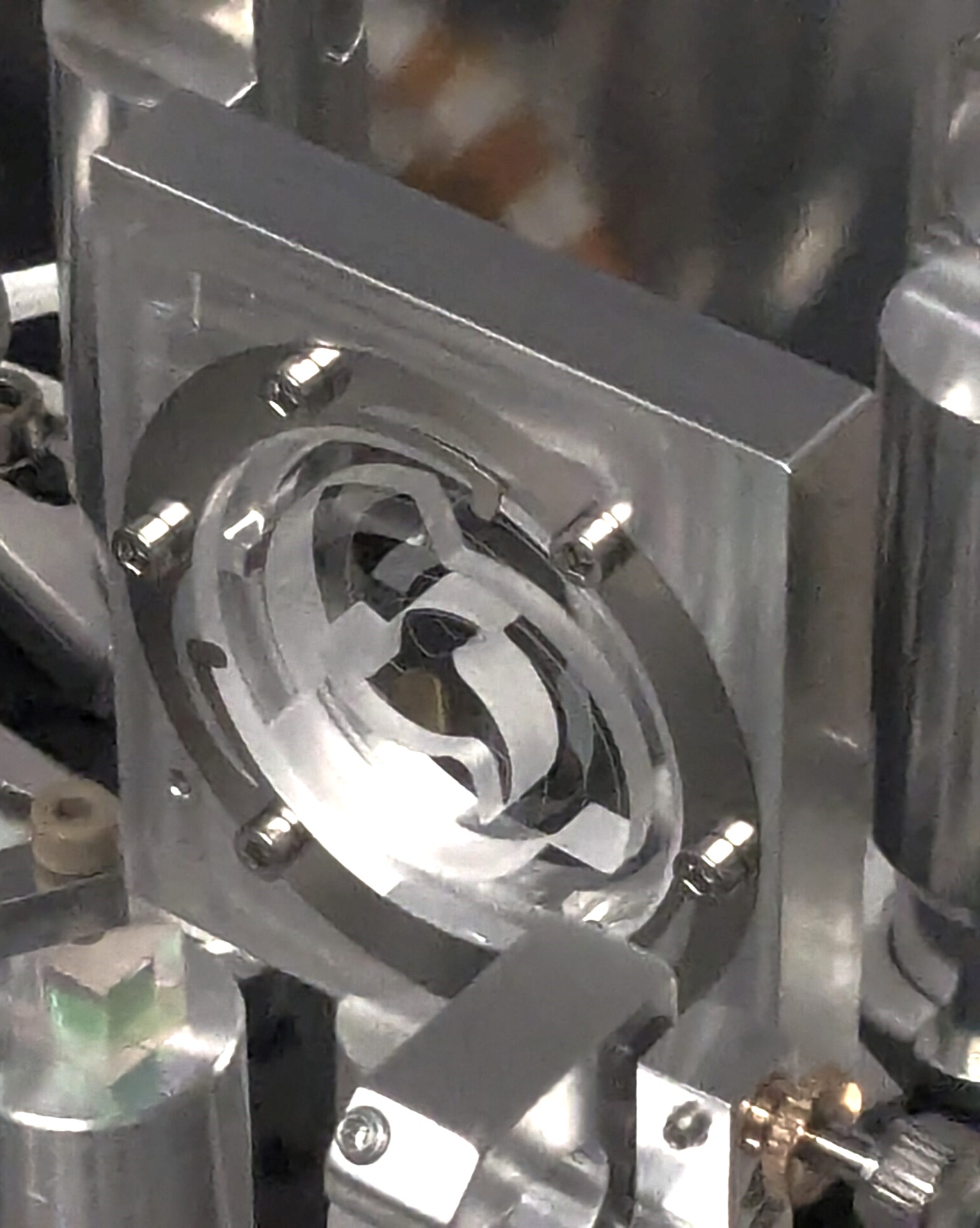}
    \hfill}\subfigure[Safety Factor Under Load in Z Direction]{\includegraphics[width=.5\textwidth]{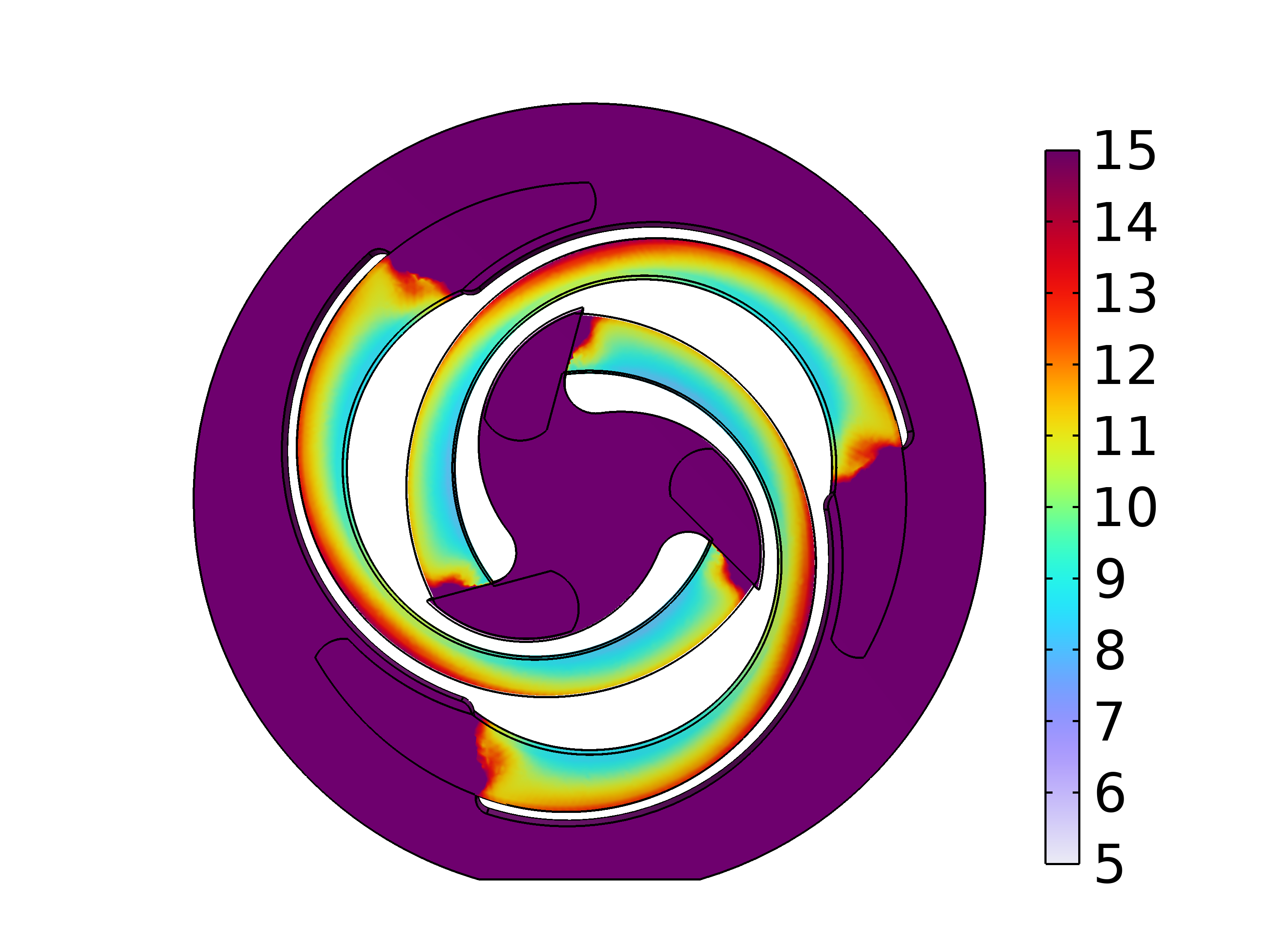}\hfill}

\caption{The design and photos of the mechanical resonator. (a) Shows a colour-coded design of the resonator. The grey is the outer frame of the resonator, which allows for mounting to the ground or table whose inertial motion is to be measured. The yellow shows the flexural bridges, which are \SI{100}{\micro\meter} thick, which act as the soft springs holding the test mass to the outer frame. The blue is the \SI{3}{\gram} test mass, the inertial mass used for encoding external motion. The reflective gold coating applied in a 5mm diameter disk to the fused silica to allow for interferometric readout is shown. (b) Shows the displacement of the test mass in the z direction during the fundamental oscillation mode, with an magnitude normalised to the maximum displacement of the piece. The face of the test mass moves along the z axis with the same displacement across its whole surface. At the same time, the outer frame remains completely still. (c) Shows a photo of the resonator in its mount with the metallic blade springs holding the outer frame in place. (d) Shows the safety factor of the resonator when under 1\,$g$ of load, as estimated by a combination of Finite Element Analysis and Equation \ref{eqn:crack}. The resonator should, therefore, be able to survive at least 5\,$g$ of load in this direction. }
\label{fig:15HZmodes}

\end{figure}
The mechanical resonator used here has been presented in a previous work~\cite{Birckigt2024}. Figure~\ref{fig:15HZmodes} shows its key features and geometry. Subtractive manufacturing techniques were reaching the limit of what could be made in terms of flexure length, so other previous work explored novel manufacturing techniques that did not scale in difficulty with the size of the piece~\cite{Birckigt2024}. The test mass and outer frame are made from one piece of glass using ultra-sonic milling; the flexures are laser etched onto a wafer. The pieces are then bonded together using plasma-activated direct bonding. The back sides of the wafers are then removed using a mechanical polish to leave the overall geometry. 
\par
Figure~\ref{fig:15HZmodes} (a) shows the parts making up the resonator in a colour-coded fashion. The outer frame (grey) is the part that is rigidly attached to the frame of reference whose inertial motion we wish to measure. It is similar to an annulus with inner parts that act as bridges for the flexure bonding. The outer ring has a 50\,mm diameter and is 6.35\,mm thick. A 50\,mm diameter piece was used as opposed to previous work with 1" optics~\cite{Carter2020a,Carter2024b} as the flexures needed to be at least 40\,mm long to survive loading from gravity as estimated by \textit{Carter et al.}~\cite{Carter2024}. The flexures were targeted to be \SI{100}{\micro\meter} thick, the inner radius is 12.2\,mm and the outer radius 15.8\,mm with the centres of the two circles being offset by 1\,mm. It was found that this lead to an approximately even stress distribution over the inner edge of the piece, minimising local stresses. The test mass has a shape that allows a large bonding area of the flexures whilst keeping some distance to the coated area. The test mass has a total weight of \SI{3}{\gram}. A reflective gold coating in a 5\ mm diameter disk in the centre was used as previous experiences with dielectric coatings have shown to degrade the Q~\cite{Carter2024thesis}. As a HoQI does not require exceptionally high reflectivity, this was not considered a limitation.
\par
The resonators were targeted to have a fundamental mode frequency of 15\,Hz. Two pieces were produced and presented in~\cite{Birckigt2024}. The first piece has a slightly higher than targeted resonance as the flexures were not thin enough. This is visible from examining the piece, where the thicker, less transparent centres of the flexures are visually apparent. A second piece was produced with 15\,Hz resonance, but was sadly damaged when mounting for the tests presented here. The mode shape of the fundamental mode is shown in Figure~\ref{fig:15HZmodes} (b). The mode is designed so that the whole face of the test mass moves in parallel. The circular symmetry of the design helps with this. The test mass moves in the Z-direction with respect to the outer frame so that its movement can easily be read out. The resonator has a collection of tip-tilt, bounce, and internal flexure modes starting from 300\,Hz and above. %Therefore, the sensor is only reliable for frequencies below this frequency. 
These frequencies are already well outside the control range of an isolation system of a gravitational wave detector, and so there are no issues when used in the controls there.
\par
Using the methods described at length in \textit{Carter et al.}~\cite{Carter2024} we estimate the contributions to the Q. The resonator has a theoretical maximum Q defined by a combination of thermoelastic damping at 20\,Hz of $Q_{\rm{TED}}=\SI{1.3e6}{}$~\cite{lifshitz2000}, surface losses with an assumed pessimistic damage layer \SI{5}{\micro\meter} deep of $Q_{\rm{Surf}}=\SI{1e6}{}$~\cite{Gretarsson1999}, and bulk losses of the Corning 7890-0F fused silica of $Q_{\rm{Bulk}}=\SI{1.1e7}{}$~\cite{Numata2002} for an upper estimate of Q at 20\,Hz of \SI{5e5}{}. This does not account for mounting losses, which were suspected to limit the Q measurements in previous work~\cite{Birckigt2024}, and additional losses from residual contamination from the release resin used in manufacturing the flexures.  The Q of the resonator was rechecked in this work approximately 9 months after the initial validation with an improved mounting structure. The new mount increased the Q to above 150,000 for the 20\,Hz mode. The new mount used a version of the three-point blade spring mount shown in~\cite{Carter2020a} upscaled for a 2" optic, and in Figure~\ref{fig:15HZmodes} (c). Previous work with these mounts on 1" samples had shown that clamping losses were low enough to measure Qs up to 600,000~\cite{Carter2024b}, indicating these pieces may now be limited by the residual contamination of the manufacturing method.
\par
The design requirement was that the resonator should survive general handling. To this end, we aimed for a large safety factor of 5 when 1\,$g$ of load is applied in the z-direction. Failure in glass materials such as this will begin from initial microcracks being expanded by stresses under load. A visual inspection of flexures suggested that there could be microcracks \SI{10}{\micro\meter} deep along both the inner and outer radius of the flexure. These would act as a typical mode I crack opening in fracture mechanics. Finite Element Analysis (FEA) was conducted to find the stresses along the flexure when under 1\,$g$ of load. The safety factor, $SF$, was calculated from the simulated stresses using the typical criterion for a plane parallel load to a microcrack in an infinite plane~\cite{1921Griffith}
\begin{equation}
    SF=\frac{K_c}{\sigma\sqrt{\pi a}}
    \label{eqn:crack}
\end{equation}
where $K_c$ is the critical fracture toughness of fused silica with a value of 0.58\,MPa/m$^{1/2}$,
$\sigma$ is the stress on the surface, and $a$ is the length of the microcracks.
This gives a heat map over the flexures as shown in Figure~\ref{fig:15HZmodes} (d), where all points meet the safety factor 5 test. %It was found that two flexures could fail before the whole piece did. 
%A new batch of resonators are in production via this method which will achieve a $f_0$ of 5\,Hz.
\subsection{Readout of Test Mass}
\begin{figure}
    \centering
    \subfigure[HoQI schematic]{\includegraphics[width=0.5\linewidth]{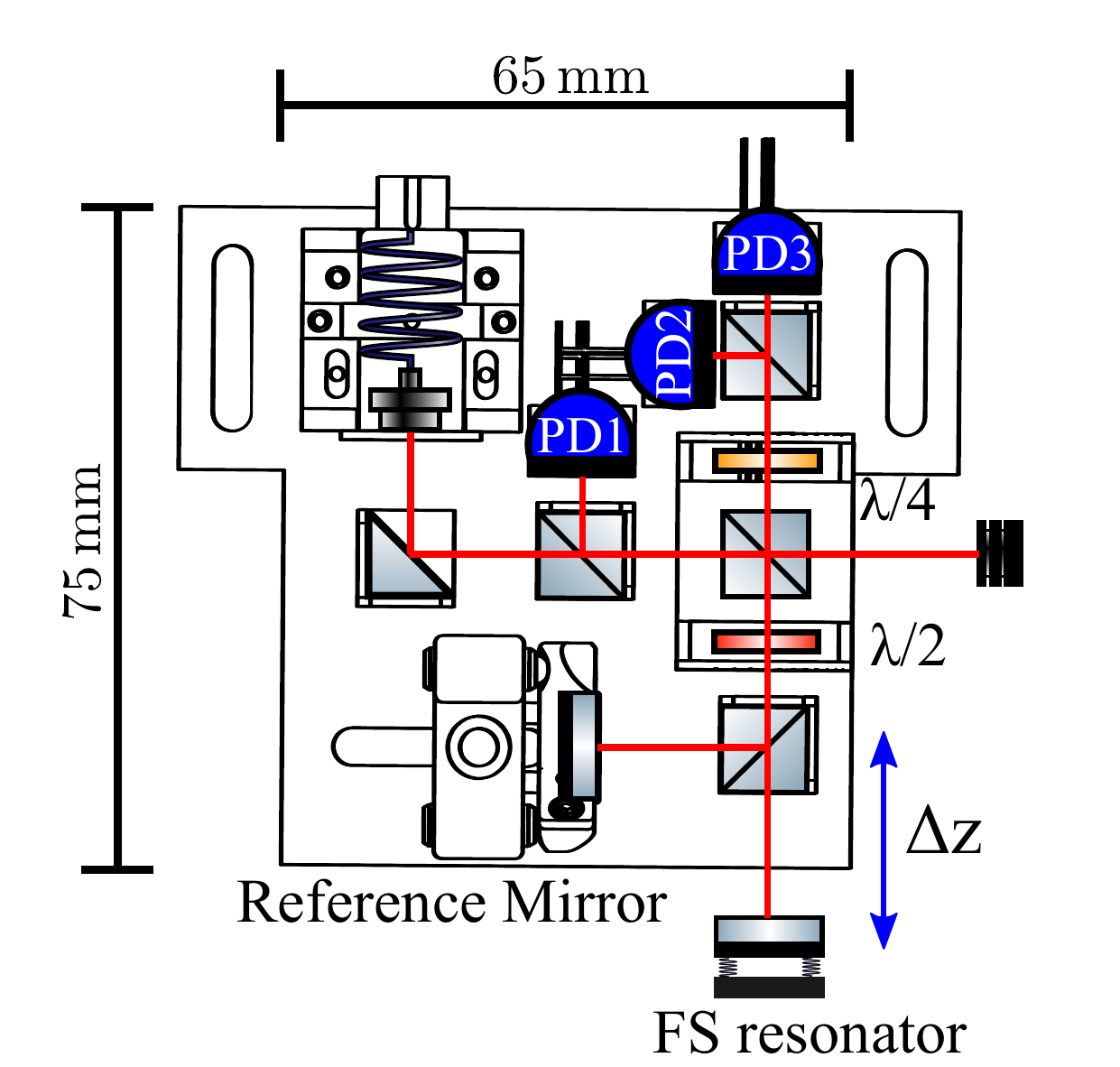}}
    \subfigure[Photo of FuSIS implementation]{\includegraphics[width=0.45\linewidth]{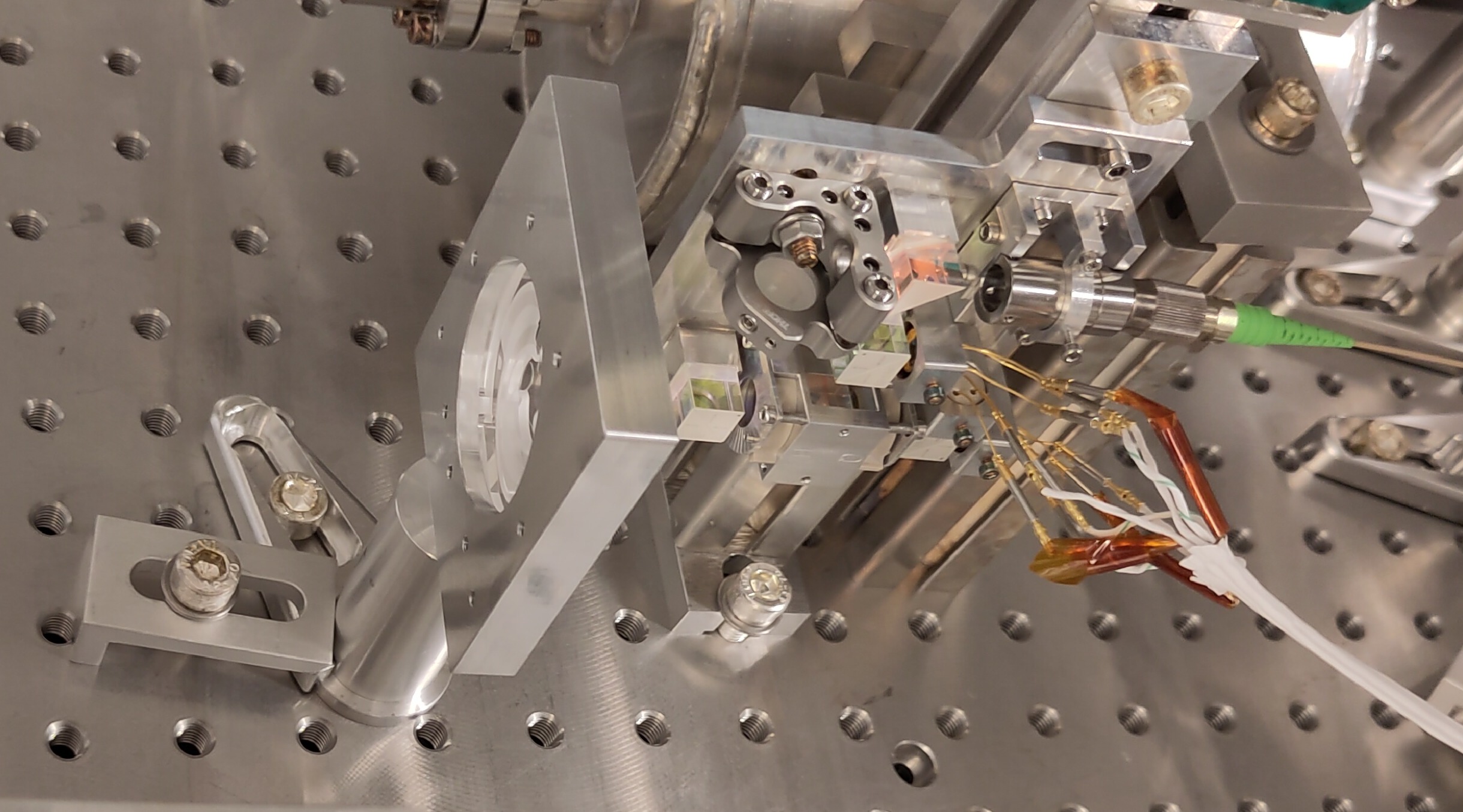}}
    \caption{A schematic diagram of the HoQI breadboard. The dimensions of the X (65\,mm) and Z (75\,mm) directions are shown for scale. The position of the fused silica (FS) resonator is labelled. In this implementation, it was not attached directly to the breadboard, but this is a simple adaption for future designs. The beam probes the Reference Mirror and resonator with differing polarisation states. These are then interfered on the photodetectors (PD) (1-3) into three complementary signals with differing phases. These can be recombined to get the optical phase difference between the reference arm and the Z position of the fused silica resonators' test mass. Full details of a HoQI's operation can be found in~\cite{Cooper2018}.   }
    \label{fig:hoqi}
\end{figure}
The HoQI used in this work was also demonstrated in a previous publication~\cite{Kranzhoff2022}. The sensing head of the HoQI was on a board shown in Figure~\ref{fig:hoqi}. Contrary to the previous publication, an Iodine stabilised NPRO laser was used to reduce potential frequency noise coupling. The light was then coupled by optical fibre from the out-of-vacuum optical preparation bench to the in-vacuum HoQI. The sensor was set up as a horizontal inertial sensor, which meant screwing the HoQI plate to a vertical post. 
\par
The trans-impedance amplifiers on this HoQI remained unchanged from previous works~\cite{Kranzhoff2022}. A LIGO-style Control and Data System (CDS) integrated into the AEI 10\,m Prototype Facility was used to record data and extract the displacement measurement from the HoQI in real-time~\cite{Gossler2010}. As such, in real-time, the three signals could be scaled with respect to each other to compensate for differing gains and optical powers in the channels. The balanced parameters were chosen and left alone throughout the measurement period of nearly two months, showing good long-term stability. All measurements presented in this work are real-time recorded displacement. 
\par
\subsection{Isolation Platform}
The AEI 10\,m Prototype has three seismically isolated platforms (Albert Einstein Institute - Seismic Attenuation System (AEI-SAS)) housed in a vacuum tank~\cite{Gossler2010,Wanner2012,Bergmann2017}. Each platform can be controlled in three translational and three rotational degrees of freedom independently. The AEI-SAS use active feedback control to further improve isolation from ground motion, using driven actuators to keep the table top inertially stable. The AEI-SAS achieves a reduction in input ground motion to tabletop motion at frequencies above 0.1\,Hz, which increases suppression by approximately two orders of magnitude at 10\,Hz~\cite{Kirchhoff2020} in the horizontal degree of freedom. The AEI-SAS has been used to verify the performance of L-4C seismometers made by Sercel~\cite{Kirchhoff2017}, a state-of-the-art kilogram scale seismometer. Other co-located witness inertial sensors are needed to verify the performance of a specific inertial sensor. By subtracting the coherent signals of each sensor from the studied sensor, the remaining signal is that sensor's self-noise. The AEI-SAS has several sensors that can measure coherent signals to subtract. There are the L-4Cs integrated on the table that are used for the active control, witness L-4Cs on the table, Suspension Platform Interferometer (SPI)~\cite{Koehlenbeck2023}, a Watt's linkage inverted pendulum sensor~\cite{Bertolini2006}, and STS-2 by Steckeisen GmbH as off-platform ground motion monitor. The L-4Cs are in all cartesian directions, distributed across the AEI-SAS, so they can be combined to give rotational degrees of freedom. 
\section{Huddle Testing in a Prototype Gravitational Wave Detector}
The combined sensor of the fused silica resonator and HoQI readout, that is, the FuSIS, was tested in a huddle test on the 10\,m Prototype's west tank AEI-SAS. The original plan was to test two such sensors side by side, but sadly, one resonator was broken while handling. This had several consequences: the broken resonator had a lower $f_0$ of 15\,Hz, so the noise performance of the 20\,Hz resonator that survived will be a factor of 2 worse at frequencies below resonance. The best noise floor we could measure would be defined by other sensors noise floors, which at frequencies below 1\,Hz is determined by the L-4C sensitivity. Above 1\ Hz, it was found that the measurements from the STS2 are sufficiently coherent to subtract them and measure a noise floor below that of the L-4C. The Watt's linkage accelerometer \cite{Bertolini2006} was found to have limited effects on the overall noise, but it did remove one specific peak which occurred at its own spring-antisping frequency. Likely, this peak was caused by back action from the moving heavy test mass of the Watt's linkage driving the table, subtly showing another advantage of gram scale test masses.
\par
The coherent subtraction used a Multi-Channel Coherent Subtraction (MCCS) routine as described in~\cite{Allen1999,spicypy}.  All the subtracted channels were inertial or displacement sensors measuring actual motion. No tertiary sensors, such as temperature sensors or laser monitors, were used, meaning the noise recorded is the inertial noise we expect from the FuSIS when used in circumstances and conditions typical of a gravitational wave detector. 
\par
15 channels in total were used as subtraction channels to find coherence. When using so many channels, random coherence in short time spans may lead to overestimating the performance. To mitigate this,  the averaging should be performed over many segments. 
In principle, two averages are required per channel used to subtract, with 15 channels used, giving a minimum of 31 averages; however, in practice, it was found that using the minimum number of averages leads to a slight underestimation of the noise floor. It was found that after 91 averages, the results had sufficiently converged to a stable estimate. To measure the very low frequencies with so many averages and sufficient sampling rates proved difficult, as any data glitch in the long-time segment would ruin a measurement. We, therefore, used a pseudo logarithmic averaging of the data whereby 31 averages were used for the lowest frequencies bins plotted (sub 10\,mHz) to over 300 averages for the higher frequencies. Our noise estimates below 10\,mHz are, therefore, slightly overestimated but are still a reasonably accurate assessment, and this is also true for our plotted L-4C noise as a comparison sensor.
\section{HoQIs as displacement readout of a quickly moving test mass}
\begin{figure}
    \centering
    \includegraphics[]{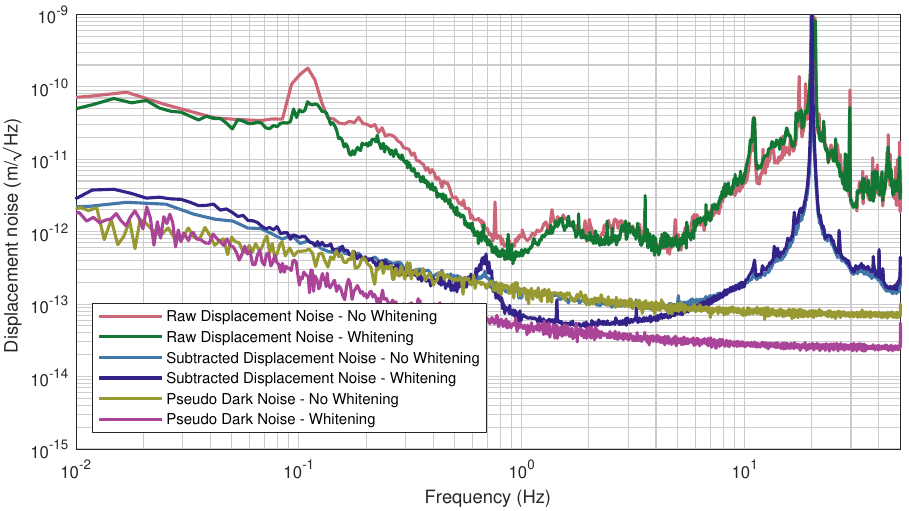}
    \caption{A graph showing the noise measuring the displacement of the test mass in the HoQI. The graph shows two settings, one with a simple flat response of the analogue electronics and one with a shaping (whitening) filter to better utilise the ADC. We had to show measurements at two different times because the measurements could not be done simultaneously with the hardware implementation. The seismic motion was slightly different at these times. The red and green lines show the raw measured noise at both times. The light and dark blue lines show the residual noise that could not be removed with reference to other sensors. The yellow and pink lines correspond to the sensor's measured pseudo dark noise.    }
    \label{fig:besthoqisDisp}
\end{figure}
Figure~\ref{fig:besthoqisDisp} shows both the raw measured displacement of the test mass with the HoQI and the residual left over when coherent subtraction is applied. The pseudo-dark noise is the noise of the electronics, including photodiodes, when plugged into the CDS while the laser is blocked. Realistic offsets are applied digitally to each channel based on measured data to create something that looks like an operating point. We see that above 0.2\,Hz the sensor was mostly limited by the dark noise. A peak at 0.7\,Hz is related to non-linear effects in the HoQI readout. An unfortunate resonance, suspected to be associated with the mounting of the HoQI plate, was present in the measurements at 20.7\,Hz. The signal from this resonance beat against the resonator's fundamental mode signal and through a non-linear downconversion, created the 0.7\ Hz resonance. This occurs because of a non-linear response in the arctangent function and residual power imbalances in the individual channels, which allows the generation of harmonics of spectral peaks. Above 7\,Hz the noise appears to follow a resonant gain; however, this is simply because only one such sensor was available for these measurements. As the sensor becomes increasingly responsive at these frequencies, its motion increases, but no other sensor is sensitive enough to subtract this. This becomes more apparent when looking at plant inverted units as we do in following sections. 
\par
As it was found that the noise limit of the HoQI was noise of the Analogue to Digital Converter (ADC) , a frequency shaping filter (or whitening filter) was integrated. To the authors' knowledge, this is the first time whitening 
has been used in a HoQI. The whitening filter is an analogue filter in each channel. These filters amplify the response in some frequencies where the motion is lower so that the ADC range is used more efficiently in all frequencies. The response of these filters can be inverted digitally to get the original signal. Components were chosen so that the amplification occurred above 0.1\,Hz to improve the response up to kHz, with a pole at 0.05\,Hz and zero at 0.15\,Hz. A modest gain of 3 was chosen as the motion on resonance was not to be amplified over the ADC range, and due to the high Q motion, this was the best that could be achieved without repeated ADC saturations. At first, after installing the whitening filter, the noise at low frequencies was increased substantially. It was found that tolerances in the passive electronic components used to make the analogue filters lead to slightly different responses of each filter, such that recombining them leads to errors. Therefore, the transfer function of each channel was measured individually, and a simple algorithm was used to fit the zero and pole of each channel uniquely. These individual parameters were then put in the de-whitening filters of each channel, and the noise improved substantially. 
The final results can be seen in Figure~\ref{fig:besthoqisDisp}, where whitening shows the expected improvement in the noise floor of the HoQI displacement noise from 0.8\,Hz to 5\,Hz. 
\par 
Below this we hit an unspecified noise floor. This noise is likely to arise due to slight misalignments of the polarisation axis of polarisation-sensitive components. We could not quantify this directly, but it was found that gently heating the fibre led to significant changes in total power in the setup and was correlated with overall phase noise. As a HoQI has significant suppression of the intensity noise of the laser, it is unclear how this noise couples into the measurement. Better control of this noise source would likely be needed to see further improvement at low frequencies. Above 2\,Hz we again see the broad resonance peak due to the higher noise level of other sensors available to validate the sensitivity. We see no evidence that it would not simply follow the ADC noise, but a second such sensor with sufficient performance at these frequencies would be needed to validate the noise in this frequency band. 
Overall, we demonstrate that despite the huge relative motion on resonance the dynamic range of a HoQI is sufficient to reach a well understood noise floor~\cite{Cooper2018,Cooper2022,Kranzhoff2022}.

\subsection{Overall Performance as an Inertial Sensor}
\begin{figure}
    \centering
  
    \subfigure[Raw Measurement]{\includegraphics[scale=.73]{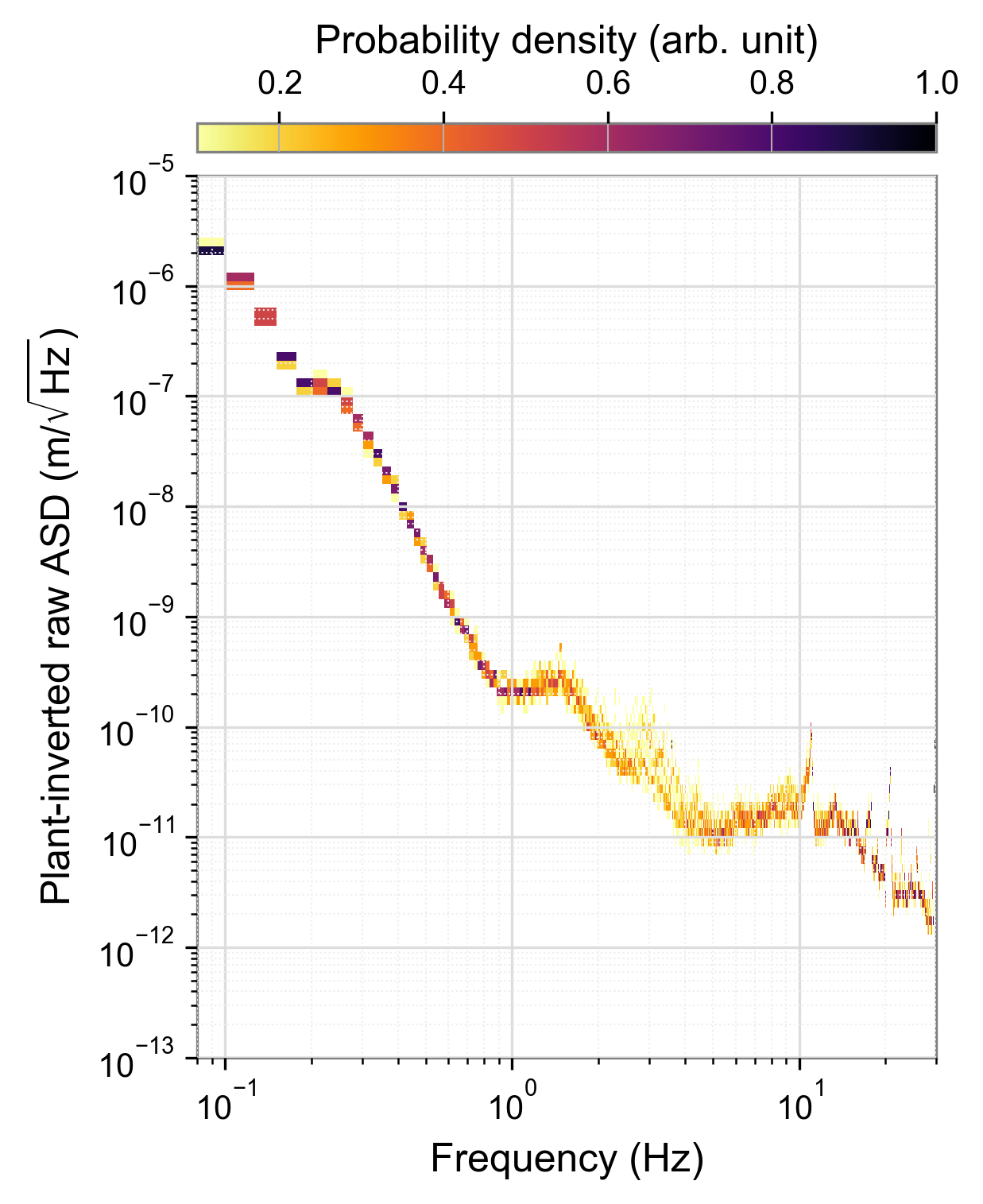}\hfill}
      \subfigure[Coherently Subtracted Residual]{\includegraphics[scale=.73]{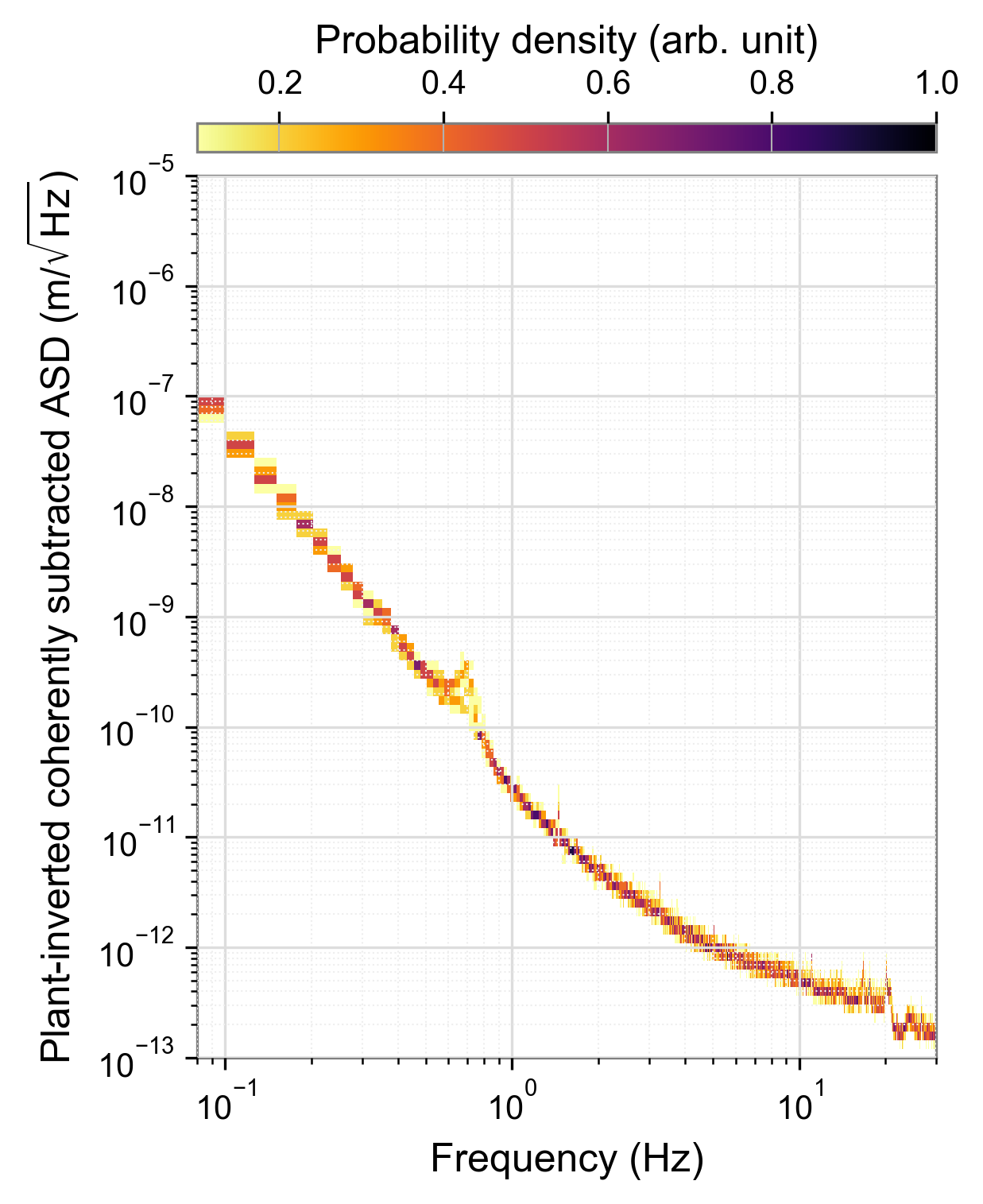}\hfill}
    \caption{A probabilistic amplitude spectral density plot  (ASD) of the raw table motion measured by the FuSIS (left), and the residual motion left after coherent subtraction (right) taken over 8 hours of measurements.}
    \label{fig:probabilityplots}
\end{figure}
To assess the FuSIS for its intended use case as a seismometer, we must invert the plant of the resonator to get the input inertial motion. This can be done using Equation~\ref{eqn:tfseismo}. 
We can also then convert between inertial displacement units and acceleration by 
\begin{equation}
    {\tilde{Z}_{\rm{g}}}(f)=\frac{1}{(2\pi f)^2}\tilde{A}_g(f)
    % \label{eqn:acctfseismo}
\end{equation}
where $\tilde{A}_g$ is the acceleration of the sensor, and $ {\tilde{Z}_{\rm{g}}}$ is the ground motion in inertial displacement.
\par
Figure~\ref{fig:probabilityplots} shows the plant inverted signal's measured raw signal and noise in a probabilistic ASD of the FuSIS. A probabilistic ASD shows the probability density of measuring a certain amplitude level throughout the length of the signal. In this case, 8 hours of quiet data over the weekend was analysed and broken into 1-hour stretches. For each 1-hour stretch, a sensor noise estimate was produced by coherent subtraction of witness sensor signals. The spread of amplitudes between these 1-hour stretches is shown with colour and line width. The lighter shade in the left panel indicates a larger range of motion amplitudes during the measurement time. These are much reduced in the right panel, which has a much darker colour, showing that the noise is broadly independent of the input signal and can be relied upon to remain constant during typical usage. The exception to this is the peak at 0.7\,Hz which shows considerable variance in height, which supports our earlier statement that this is connected to down mixing of the resonance at 20.7\,Hz. The plot is produced using the same frequency-domain averaging method as used for Figure~\ref{fig:besthoqisDisp} and described in~\cite{Allen1999}, in this case using 91 averages for each 1 hour stretch. The algorithm is implemented in the \texttt{Spicypy} python package~\cite{spicypy}, which is based on the \texttt{GWpy} package~\cite{gwpy}.
\subsection{Comparison with other Sensors}
\begin{figure}
    \centering
    \includegraphics{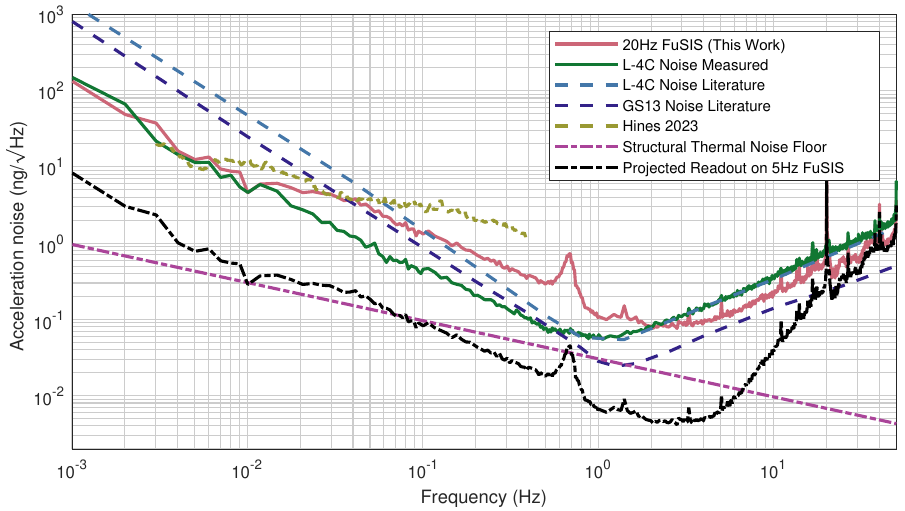}
    \caption{ Acceleration noise of the FuSIS presented in this paper and several state-of-the-art large-scale inertial sensors. The pink line shows the measured acceleration noise of our 20\,Hz resonator with a HoQI readout.  The dashed L-4C and GS13 lines correspond to noise performances presented in previous literature, while the solid green L-4C line corresponds to our measured L-4C noise with custom-built amplifiers. Yellow dashed shows the previously best-reported noise of a fused silica sensor. The structural thermal noise is the thermal noise floor of the 20\,Hz resonator estimated by Equation~\ref{eqn:strucNoise}. The black dashed line corresponds to the case the noise of this readout was put on a FuSIS with a $f_0=5$\,Hz such as those shown in~\cite{Hines2023} or as will be built in the next year for future iterations of this work. }
    \label{fig:AccComp}
\end{figure}
Figure~\ref{fig:AccComp} compares the noises of this sensor along with several other state-of-the-art inertial sensors used today. 
\par
The coloured dashed lines in Figure~\ref{fig:AccComp} show sensor noises reported in literature~\cite{Hines2023,Carter2024,Matichard2015a,Matichard2015b} of the L-4C and GS13. However, the noise of the L-4C with well-optimised amplifiers has been shown to be better at low frequencies~\cite{Kirchhoff2020}, a result which we reproduce in this work by performing coherent subtraction using the L-4C as the base sensor, which gives the green line. The yellow dashed line shows the literature's previous best-reported fused silica sensor. The sensor presented in this work (pink line) shows better sensitivity from 1\,mHz to 20\,Hz, whereas a previous iteration of this sensor has shown better sensitivity~\cite{Carter2024b} above this frequency band. At frequencies below 10\,mHz, our assessment of the FuSIS's performance is limited by sensors available for comparison. The FuSIS performs at a comparable level to other state-of-the-art inertial sensors in this band, despite being a fraction of the size, with the added benefit of being self-calibrating~\cite{Gerberding2015} and linear in frequency response due to the mechanical behaviour of the resonator.
\par
Furthermore, as the test mass is only 3\,g and not magnetic, we would not get the complicated back action induced by the motion of otherwise substantial payloads or the corresponding magnetic field fluctuations. The 15\,Hz resonator shown in a previous publication about the manufacturing~\cite{Birckigt2024} was damaged. If the 20\ Hz resonator was the damaged one instead, we likely would have gained a factor of 2 in performance outright at frequencies below $f_0$ and, therefore, have matched the L-4C across the entire band, outside of the down-converted peak.
\section{Conclusions and Outlook}
This paper has demonstrated that Homodyne Quadrature Interferometers (HoQIs) are an excellent choice for reading out the motion of test masses in high Q inertial sensors. The HoQI's performance does not degrade despite the large dynamic range of the readout. We achieve a noise level in a HoQI comparable with the best reported in the literature to date. Furthermore, we show that shaping filters can be used without sacrificing sensitivity at any frequency to better utilise ADC range. That this works even when measuring high Q induced motion shows that the technique is broadly applicable to any HoQI application. 
\par
When used in a fused silica resonator based inertial sensor (FuSIS), it performs comparably to kilogram-scale state-of-the-art sensors. This makes them an excellent candidate for any vacuum application with inertial sensing requirements because of their compact size, self-calibration, and inherent vacuum compatibility. A relatively simple change to silicon as material would make such sensors suited for cryogenic applications. The HoQIs have the advantage of having no active control and are capable of measuring motions over several optical fringes, truly making FuSIS ``plug and play". The sensors can work for long periods without needing adjustment or worrying about the test masses getting stuck. A quick impulse applied to the sensor can automatically verify or update any calibrations. All these combine to make the FuSIS design practical and valuable for a wide range of applications, with seismic isolation of gravitational wave detectors just one example.
\par
As previously stated, the resonators used so far were proof-of-principle pieces for a novel manufacturing method~\cite{Birckigt2024}. As an outlook, the goal now is to produce fused silica resonators with much lower $f_0$ (targeting 5\,Hz). Other groups have demonstrated resonators with a $f_0$ of 5\,Hz, suggesting that this is possible~\cite{Hines2020,Hines2023}. If a HoQI were used with such an envisioned resonator, we would likely get a noise as shown by the black dashed line in Figure~\ref{fig:AccComp}. Such a sensor would be an excellent candidate for use in feedback isolation systems, as it would be sensitive over the entire control band, removing the need for sensor blending of Cartesian inertial sensors~\cite{Matichard2015a,Matichard2015b,Carter2020}. At the same time, at the critical frequencies above 1\,Hz it would offer unparalleled sensitivity. This could be a crucial technology for the next generation of gravitational wave detectors, such as the Einstein Telescope~\cite{Punturo2010}, to meet their stringent requirements on seismic isolation. 

\ack
JJC and PB acknowledge funding in the framework of the Max-Planck-Fraunhofer cooperation project ``Glass technologies for the {Einstein} Telescope (GT4ET)''.
Authors JL, AB, SAL, MC, GC, FK, HL, PS, ST, JvW, DSW acknowledge funding by the Deutsche Forschungsgemeinschaft (DFG, German Research Foundation) under Germany’s Excellence Strategy – EXC-2123 QuantumFrontiers – 390837967.
The authors would like to thank David Hoyland from the University of Birmingham for providing schematics for the HoQI readout electronics. We would also like to thank the advice and discussion from Conor Mow-Lowry, Alex Mitchell, and Gerhard Heinzel.
\section*{References}
\bibliographystyle{iopart-num}
\input{ref}
\end{document}

%% file: ref.tex
\providecommand{\newblock}{}